\documentclass[nonacm,screen]{acmart}
\thanks{CSCW 2021 Workshop -- Investigating and Mitigating Biases in Crowdsourced Data, October 23, 2021, Virtual.\\ \copyright 2021 Copyright held by the author(s).}
\authorsaddresses{}
\AtBeginDocument{%
  \providecommand\BibTeX{{%
    \normalfont B\kern-0.5em{\scshape i\kern-0.25em b}\kern-0.8em\TeX}}}

\setcopyright{acmcopyright}
\copyrightyear{2021}
\acmYear{2021}
\acmDOI{10.1145/1122445.1122456}

\acmConference[CSCW Workshop (Investigating and Mitigating Biases in Crowdsourced Data)]{CSCW 2021 Workshop - Investigating and Mitigating Biases in Crowdsourced Data}{23 October 2021}{Virtual}
\acmBooktitle{CSCW 2021 Workshop - Investigating and Mitigating Biases in Crowdsourced Data, 23 October 2021, Virtual}
\acmPrice{15.00}
\acmISBN{978-1-4503-XXXX-X/18/06}

\begin{document}

\title[Demographic Biases in KOLs]{Demographic Biases of Crowd Workers in Key Opinion Leaders Finding}

\author{Hossein A.~Rahmani}
\affiliation{%
  \institution{University College London}
  \city{London}
  \country{United Kingdom}}
\email{h.rahmani@ucl.ac.uk}

\author{Jie Yang}
\affiliation{%
  \institution{Delft University of Technology}
  \city{Delft}
  \country{The Netherlands}
}
\email{j.yang-3@tudelft.nl}

\renewcommand{\shortauthors}{Hossein A.~Rahmani and Jie Yang}

\begin{abstract}
Key Opinion Leaders (KOLs) are people that have a strong influence and their opinions are listened to by people when making important decisions. Crowdsourcing provides an efficient and cost-effective means to gather data for the KOL finding task. However, data collected through crowdsourcing is affected by the inherent demographic biases of crowd workers. To avoid such demographic biases, we need to measure how biased each crowd worker is. In this paper, we propose a simple yet effective approach based on demographic information of candidate KOLs and their counterfactual value. We argue that it is effectiveness because of the extra information that we can consider together with labeled data to curate a less biased dataset.
\end{abstract}


%   <concept>
%       <concept_id>10002951.10003317.10003347.10003354</concept_id>
%       <concept_desc>Information systems~Expert search</concept_desc>
%       <concept_significance>500</concept_significance>
%       </concept>
%  </ccs2012>

% \ccsdesc[500]{Information systems~Expert search}

%   <concept>
%       <concept_id>10003120.10003130.10003134</concept_id>
%       <concept_desc>Human-centered computing~Collaborative and social computing design and evaluation methods</concept_desc>
%       <concept_significance>500</concept_significance>
%       </concept>
%  </ccs2012>

% \ccsdesc[500]{Human-centered computing~Collaborative and social computing design and evaluation methods}


\maketitle

\section{Introduction}
Key Opinion Leaders (KOLs) are people that have such a strong social and professional status that their recommendations and opinions are listened to when making important decisions. For instance, in the field of medical and health informatics, KOLs are the people who can influence public opinion and lead the medical community through their research papers, clinical practices, and early acceptance of new technologies. Traditionally, consulting companies provide services for identifying KOLs by conducting user surveys. The problems of these solutions are that they use only a limited number of information resources and focus on a small number of involved clients. Therefore, they are not very effective in the real scenarios as well as sensitive domains. Existing studies \cite{li2013improved, xu2010medex} address these problems using Machine Learning (ML) approaches that are scalable and are able to deal with a large number of candidate KOLs. However, ML approaches require large amount of labeled training data. The datasets are hand-labeled by people who are domain experts and usually very hard to gather: finding KOLs is a time-consuming and typically difficult process even for domain experts. Consequently, training models based on such datasets makes them highly dependent and limited to expert labels. By allowing to reach to large number of online crowds, crowdsourcing has recently become one of the most promising approaches in collecting data for training ML models for different tasks such as {\itshape political ideology detection}, {\itshape detecting biased statements}, and {\itshape finding social influencer} \cite{gadirajucan2020crowd,parshotam2013crowd,hube2018detecting,iyyer2014political,arous2020opencrowd}. However, in many tasks, like KOL mapping, the annotation data are usually affected by the biases of the crowd workers.

In this paper, we consider a crowdsourcing task where the crowd workers are asked to name as many as KOLs as possible in a specific domain. We propose an approach to measure how biased a crowd worker is, through which we can mitigate worker biases and clean the collected data from biased crowd workers.

\section{Related Work}
Recent works have explored the mitigating crowd worker biases, for example, \citet{hube2019understanding} focused on the subjective task and tried to understand the influence of worker's preferences on their performance. To do this, they examine the annotations of crowd workers on different topics to see the effect of worker's opinions on their annotations. Their findings show that crowd workers with strong opinions produce biased annotations. The proposed approach is promising to mitigate such bias and can improve the quality of the data collected. \citet{chakraborty2017makes} analyzed the demographics of people who suggest the recommendation of contents to understand the demographic distribution of content promoters in social networks. This distribution can show whether these people are representative of the social network population or there is a bias to the groups of people. To this end, they collect extensive data from Twitter of trending topics and study the demographic biases of trends. Their analysis indicates that a large part of the demographic information of crowds who promoted the trends is significantly different from the overall Twitter population. In \cite{hube2018limitbias}, the authors extensively analyzed the effect of crowd worker's opinions on the quality of the annotated data. They proposed an approach that relies on the labels of the statements and the worker's personal opinion on each statement's topic. Using this additional information, they are able to measure how biased a crowd worker is and how they can mitigate the measured bias. Raykar et al., in \cite{raykar2009supervised}, proposed an approach based on the combination of labels provided by different types of crowd workers, i.e., experts and beginners. Therefore, to acquire the final labels of the task they can evaluate the different labels from both experts and beginners then give an estimate of the actual labels. Few researchers \cite{karger2011iterative, liu2012variational} have addressed the problem of crowd workers' biases by assuming different experts between the crowd workers and based on that they proposed the label aggregation models. These approaches usually improve the collected labels by the majority voting among the workers. However, the idea is suitable when there is no agreement between the workers; in subjective tasks such as KOLs, there may be biases also with complete agreement labels due to the varying ideological backgrounds of workers.	

\section{Proposed Approach}
We propose an approach for collecting data and measuring crowd worker biases for the task of mapping candidate/potential KOLs as either KOL or non-KOL. Our approach can be used for other similar tasks.

Our approach automates the finding and suggesting of potential candidate KOLs. To achieve this, we will prepare a crawling module that collects information using different APIs and scarping different sources. In this study, we target two different aspects of KOLs related to their professionalism in their topics (i.e., \verb|scientific aspect|) as well as their socialites' expertise in organizing events and conferences (i.e., \verb|social aspect|). In particular, we consider \textit{Google Scholar}\footnote{\url{https://scholar.google.com/}}, \textit{PubMed}\footnote{\url{https://pubmed.ncbi.nlm.nih.gov/}}, and \textit{ClinicalTrials.gov}\footnote{\url{https://clinicaltrials.gov/}}. But our crawling module is not limited only to these sources and it is able to be applied to different information sources on various domains. In the next step, we ask crowd workers to suggest as many candidate KOLs as possible. Here, the KOL is about ``influence'', and crowds are the target who are directly addressed. The crawling module provides useful information which helps crowd workers to carefully indicate the candidate KOLs. For example, in the category of scientific information, the number of citations of the potential KOLs can be a good parameter to evaluate the quality of candidate KOLs. To this end, we will present the set of collected features representing a candidate KOL, namely, demographic, scientific, and social information to a crowd worker. The KOL mapping task is to predict the likelihood of a candidate KOL to be a potential KOL on a rating scale based on the collected features. Then, we will ask each crowd worker to label $k$ out of $N$ candidate KOLs where $N$ is the number of all candidate KOLs. In this step, to consider the crowd worker bias, we generate the counterfactual of the features for those $k$ candidate KOLs.

In this study, what we are considering is a simple yet effective class of counterfactual which can be generated by changing the value of demographic information such as age, sex, and race of candidate KOLS. For example, if we want to deal with gender bias what can we do is generating counterfactual information of candidate KOLs when we change their sex attributes. As shown in Eq.~\ref{eq:worker_bias}, we compute biases of crowd workers using the mean absolute difference of rating score provided for all $k$ pairs of the main candidate KOL and the counterfactual candidate KOL as follows:

\begin{equation}
    \text { WorkerBias }=\frac{1}{k} \sum_{i=1}^{k}\left|\mathcal{MK}_{i}-\mathcal{CK}_{i}\right|
    \label{eq:worker_bias}
\end{equation}

\noindent where the $\mathcal{MK}_i$ and $\mathcal{CK}_i$ are the rating score for the main and counterfactual candidate KOL, respectively. Future work will concentrate on extend Eq.~\ref{eq:worker_bias} to a weighted biased crowd workers that consider all feature categories, i.e., demographic, scientific, and social aspects. In this case, crowd workers will assign rating scores for each dimension and we compute the bias score as follows:

\begin{equation}
    \text{WorkerBias}=\alpha(\frac{1}{k}\sum_{i=1}^{k}\left|\mathcal{MK}_{di}-\mathcal{CK}_{di}\right|) + \beta(\frac{1}{k}\sum_{i=1}^{k}\left|\mathcal{MK}_{ci}-\mathcal{CK}_{ci}\right|) + (1 - \alpha - \beta)(\frac{1}{k}\sum_{i=1}^{k}\left|\mathcal{MK}_{si}-\mathcal{CK}_{si}\right|)
    \label{eq:extend_bias}
\end{equation}

\noindent where the $\mathcal{MK}_{di}$ and $\mathcal{CK}_{di}$ are the rating score related to the demographic aspect, $\mathcal{MK}_{ci}$ and $\mathcal{CK}_{ci}$ are correspond to the scientific aspect, and $\mathcal{MK}_{si}$ and $\mathcal{CK}_{si}$ are related to the social aspect.

The final score indicates how biased is a crowd worker; the lower the \verb|WorkerBias| score, the lower unbiased behavior of crowd worker, and in contrast, the higher values of the \verb|WorkerBias| score shows a more biased behavior of crowd workers. Therefore, we can use this information of crowd workers bias in conjunction with the crowd worker responses to collect fairer labels and achieve a better dataset. For instance, we can define a threshold based on the biased scores of crowd workers and filter out the labels from those crowd workers whose biased score is beyond the threshold. There may be an issue with generating counterfactual sensitive attributes when a crowd worker relates a previously rated candidate KOL with its counterfactual. This makes a problem to understand is really a crowd worker biased when the crowd worker realized she rated a very similar candidate KOL just before. To address this issue we can consider several solutions: (1) we can play with the order of the candidate KOL information which will be presented to the crowd worker, we should place the candidate information far from each other; (2) we can add noise to some features, for instance, we can change the age of the candidate KOL; (3) we can change the irrelevant and unimportant attributes that those attributes will not have any effects on the crowd worker's rate, for example, first name, last name, email or phone number, etc. 

\section{Conclusion and Future Work}
In this position paper, we propose a simple yet effective method to measure the demographic biases of crowd workers using a counterfactual approach. Although introduce this approach on the key opinion leaders finding problem, our proposed method can be applied to any social computing problem, when crowd workers classify data based on the social or demographic information. In our future work, we first plan to evaluate this approach using an empirical study by comparing the dataset obtained using our approach and other reported results. Next, we want to extend this approach with a polynomial regression approach when we can consider different weights for each attribute. Finally, we plan to explore how the existing methods fare against different fairness metrics.

\bibliographystyle{ACM-Reference-Format}
\bibliography{sample-base}

% \appendix

\end{document}